\documentclass[aps,pra,twocolumn,superscriptaddress,letterpaper]{revtex4-1}
\usepackage{graphicx}
\usepackage{amsmath}
\usepackage{esint}
\usepackage{verbatim}
\usepackage{color}
\usepackage{SIunits}
\usepackage{hyperref}

\newcommand{\nbar}{\langle n \rangle}
\newcommand{\nbath}{n_{\text{b}}}

\newcommand{\ncavO}{n_{\text{c}}}
\newcommand{\gzeroO}{g_{\text{0}}}

\newcommand{\kappai}{\kappa_{\text{i}}}
\newcommand{\kappae}{\kappa_{\text{e}}}
\newcommand{\gammaiO}{\gamma_{\text{i}}}

\newcommand{\omegacO}{\omega_{\text{c}}}
\newcommand{\omegamO}{\omega_{\text{m}}}
\newcommand{\gammaOMO}{\gamma_{\text{OM}}}
\newcommand{\QoO}{Q_{\text{c}}}
\newcommand{\QmO}{Q_{\text{m}}}

\newcommand{\omegalO}{\omega_{\text{l}}}
\newcommand{\ahat}{\hat{a}}
\newcommand{\adag}{\hat{a}^{\dagger}}
\newcommand{\bhat}{\hat{b}}
\newcommand{\bdag}{\hat{b}^{\dagger}}
\newcommand{\gTwoTau}{g^{(2)}(\tau)}
\newcommand{\gTwoZero}{g^{(2)}(0)}
\newcommand{\nNEP}{n_\text{NEP}}
\newcommand{\nPump}{n_\text{pump}}
\newcommand{\nDark}{n_\text{dark}}
\newcommand{\gammaSB}{\Gamma_\text{SB,0}}
\newcommand{\gammaPump}{\Gamma_\text{pump}}
\newcommand{\gammaDark}{\Gamma_\text{dark}}
\newcommand{\etaSPD}{\eta_\text{SPD}}
\newcommand{\Ib}{I_\text{b}}

\newcommand{\Isw}{I_\text{sw}}
\newcommand{\Tc}{T_\text{c}}
\newcommand{\Vd}{V_\text{d}}

\begin{document}

\title{Phonon counting and intensity interferometry of a nanomechanical resonator}

\author{Justin D.\ Cohen}
\thanks{These authors contributed equally to this work.}
\author{Se\'{a}n M.\ Meenehan}
\thanks{These authors contributed equally to this work.}
\author{Gregory S. MacCabe}
\affiliation{Institute for Quantum Information and Matter and Thomas J. Watson, Sr., Laboratory of Applied Physics, California Institute of Technology, Pasadena, CA 91125, USA}
\author{Simon Gr\"{o}blacher}
\affiliation{Institute for Quantum Information and Matter and Thomas J. Watson, Sr., Laboratory of Applied Physics, California Institute of Technology, Pasadena, CA 91125, USA}
\affiliation{Vienna Center for Quantum Science and Technology (VCQ), Faculty of Physics, University of Vienna, A-1090 Wien, Austria}
\author{Amir H. Safavi-Naeini}
\affiliation{Institute for Quantum Information and Matter and Thomas J. Watson, Sr., Laboratory of Applied Physics, California Institute of Technology, Pasadena, CA 91125, USA}
\affiliation{Edward L. Ginzton Laboratory, Stanford University, Stanford, CA 94305, USA}
\author{Francesco Marsili}
\author{Matthew D. Shaw}
\affiliation{Jet Propulsion Laboratory, 4800 Oak Grove Dr., Pasadena, CA 91109, USA}
\author{Oskar Painter}
\email{opainter@caltech.edu}
\affiliation{Institute for Quantum Information and Matter and Thomas J. Watson, Sr., Laboratory of Applied Physics, California Institute of Technology, Pasadena, CA 91125, USA}

\date{\today}
\begin{abstract}
Using an optical probe along with single photon detection we have performed effective phonon counting measurements of the acoustic emission and absorption processes in a nanomechanical resonator.  Applying these measurements in a Hanbury Brown and Twiss set-up, phonon correlations of the nanomechanical resonator are explored from below to above threshold of a parametric instability leading to self-oscillation of the resonator.  Discussion of the results in terms of a ``phonon laser'', and analysis of the sensitivity of the phonon counting technique are presented.   
\end{abstract}
\pacs{}
\maketitle

In optics, the ability to measure individual quanta of light (photons) enables a great many applications, ranging from dynamic imaging within living organisms~\cite{Hoover2013}, to secure quantum communication~\cite{Hadfield2009}. Landmark photon counting experiments, such as the intensity interferometry performed by Hanbury Brown and Twiss (HBT)~\cite{HanburyBrown1956b} to measure the angular width of visible stars, have played a critical role in our understanding of the full quantum nature of light~\cite{Glauber1963c}.  In this work we use an optical probe and single photon detection to study the acoustic emission and absorption processes in a silicon nanomechanical resonator, and perform a similar HBT measurement to measure correlations in the emitted phonons as the resonator undergoes a parametric instability formally equivalent to that of a laser~\cite{Grudinin2010}.  Owing to the cavity-enhanced coupling of light with mechanics, this effective phonon counting technique has a noise equivalent phonon sensitivity of $\nNEP = 0.89 \pm 0.05$ at an intracavity photon number of order unity.  With straightforward improvements to this method, a variety of quantum state engineering tasks employing mesoscopic mechanical resonators will be enabled~\cite{Vanner2013}, including the generation and heralding of single phonon Fock states~\cite{Galland2014} and the quantum entanglement of remote mechanical elements~\cite{Borkje2011,Lee2011}.           

Mechanical resonators are currently employed in myriad technologies, from precision navigation systems~\cite{Rozelle2009} to personal electronics~\cite{Weigel2002}.  As with objects at the atomic scale, the properties of macroscopic mechanical resonators are governed by the laws of quantum mechanics, providing fundamental quantum limits to the sensitivity of mechanical sensors and transducers.  Current research in the area of cavity-optomechanics seeks to explore the quantum mechanical properties of mechanical systems ranging in size from kilogram mass mirrors to nanoscale membranes~\cite{Aspelmeyer2013}, and to develop technologies for precision sensing~\cite{Suh2014} and quantum information processing~\cite{Stannigel2010,Safavi-Naeini2011a,Stannigel2012,Bochmann2013,Andrews2014}.  Measurement of the properties of mechanical systems in the quantum regime typically involve heterodyne detection of a coupled optical or electrical field, yielding a continuous signal proportional to the mechanical displacement amplitude~\cite{Clerk2010}.  An alternative method, particularly suited to optical read-out, is to utilize photon counting as a means to probe the quantum dynamics of the coupled opto-mechanical system~\cite{Qian2012,Kronwald2012}.  

Photon counting can be readily adapted to study intensity correlations in an optical field, and has been used not only in the astronomical HBT studies of thermal light, but also in early studies of the photon statistics of laser light and single-atom fluorescence~\cite{Glauber1963c,Kimble1977}.  In the field of photon-correlation spectroscopy, such intensity interferometry techniques have found widespread application in the measurement of particle and molecular motion in materials~\cite{Pike2010}.  More recently, photon counting of Raman scattering events in diamond has been used to herald and verify the quantum entanglement of a THz phonon shared between two separate bulk diamond crystals~\cite{Lee2011}.  In the case of engineered cavity-optomechanical systems, much longer phonon coherence times are attainable, albeit at lower mechanical frequencies (MHz-GHz) which limit the temperature of operation and the optical power handling capability of such structures.  Quantum optical schemes for manipulation of the quantum state of motion in cavity-optomechanical systems thus rely on a large per-phonon scattering rate and efficient detection of scattering events.  In this work we embed a high-$Q$, GHz-frequency, mechanical resonator inside an optical nanocavity, greatly enhancing the phonon-photon coupling rate and channeling optical scattering into a preferred optical mode for collection, both of which combine to yield a sub-phonon-level counting sensitivity of the intracavity mechanical resonator occupancy.

%In each of these cases there is a different tendency of photons: they can either bunch together (thermal light)~\cite{Morgan1966}, remain uncorrelated in time (laser light)~\cite{Glauber1963c}, or anti-bunch (single atom resonance fluorescence)~\cite{Kimble1977} with some finite time separation.

%Modern experiments utilize photon counting to generate and herald quantum states of light~\cite{Hong1986}, to entangle remote quantum systems~\cite{Chou2005}, and for linear optics quantum computing~\cite{Knill2001}.  

%Early on it was recognized that similar intensity interferometry techniques could be used to perform high resolution spectroscopy photon counting could also be used to study phonons~\cite{Fetter1965}, and similarly i

%In the case of engineered cavity-optomechanical systems, with their typically limited optical power handling capability and thermal noise sensitivity, the combination of large per-phonon scattering rate and efficient scattering detection are crucial technical elements required of any scheme proposing to             

\begin{figure}[btp]
\begin{center}
\includegraphics[width=\columnwidth]{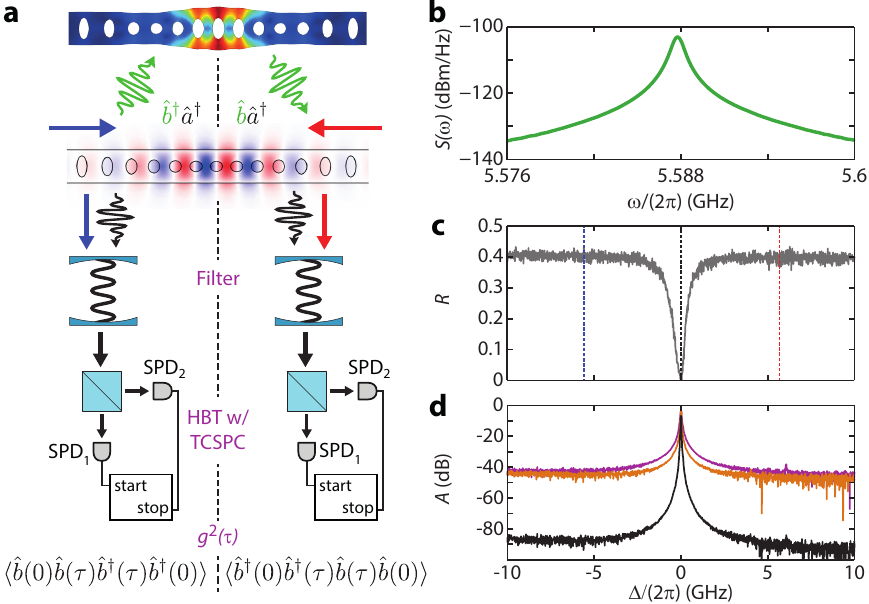}
\caption{\textbf{a}, Schematic of the phonon counting measurement. The finite element method simulations depict the displacement field of the acoustic resonance (top) and the electric field of the optical resonance. Pump light at optical detuning $\Delta = -\omegamO$ ($\Delta = \omegamO$) is indicated by the blue (red) arrows and corresponds to measurement of phonon emission (absorption), while optically resonant light is represented by black arrows. The optical reflection is filtered to reject the pump, then detected in a Hanbury Brown and Twiss (HBT) set-up using two superconducting single photon detectors (SPDs), with outputs used as start/stop pulses in a time-correlated single photon counting module. Left side of the schematic shows the resulting anti-normal-ordered phonon correlation function for blue detuned pumping ($\Delta = -\omegamO$), while right side of the schematic shows red detuned pumping ($\Delta = \omegamO$) which yields a normal-ordered correlation function. \textbf{b}, Measured power spectral density $S(\omega)$ of the acoustic resonance. \textbf{c}, Normalized optical cavity reflection spectrum $R$. Pump detunings of $\Delta = \pm \omegamO/(2\pi) = \pm 5.6$~GHz are indicated by the red and blue dashed lines, respectively. \textbf{d}, Transmission spectrum of the first (purple) and second (orange) optical filter, with total filter transmission plotted in black.} \label{fig:setup}
\end{center}
\end{figure}

A conceptual schematic of the phonon counting experiment presented here is shown in Fig.~\ref{fig:setup}a. The device studied in this work consists of a patterned silicon nanobeam which forms an optomechanical crystal (OMC)~\cite{Chan2011,Chan2012} able to co-localize acoustic and optical resonances at frequencies $\omegamO$ and $\omegacO$, respectively. Finite-element method simulations of the acoustic and optical resonances are shown at the top of Fig.~\ref{fig:setup}a. The Hamiltonian describing the interaction between the acoustic and optical modes is given by $\hat{H}_{\text{int}} = \hbar \gzeroO \adag \ahat (\bhat + \bdag)$, where $\ahat$ ($\bhat$) is the annihilation operator for the optical (acoustic) mode, and $\gzeroO$ is the optomechanical coupling rate, physically representing the optical frequency shift due to the zero-point motion of the acoustic resonator. This interaction modulates an optical cavity drive at frequency $\omegalO$ to produce sidebands at frequencies $\omegalO \pm \omegamO$, analogous to the anti-Stokes and Stokes sidebands in Raman scattering and corresponding to phonon absorption or emission, respectively. For a system in the resolved sideband limit, where $\omegamO \gg \kappa$ ($\kappa$ is the linewidth of the optical resonance), the density of states of the optical cavity can be used to resonantly enhance either scattering process for an appropriately detuned pump. In particular, applying a large coherent pump red (blue) detuned from the optical cavity resonance by $\Delta = \omegacO - \omegalO = \omegamO$ ($\Delta = -\omegamO$) results in an effective interaction Hamiltonian of the form $\hat{H}_{\text{int}} = \hbar G ( \adag \bhat + \ahat \bdag)$ ($\hat{H}_{\text{int}} = \hbar G (\ahat \bhat + \adag \bdag)$), where $G = \gzeroO \sqrt{\ncavO}$ is the parametrically enhanced optomechanical coupling rate ($\ncavO$ is the intracavity photon number at frequency $\omegalO$ due to the pump laser). In this case, the output field annihilation operator $\ahat_{\text{out}}$ can be shown to consist of a coherent component at frequency $\omegalO$ as well as a component at frequency $\omegacO$ which is proportional to $\bhat$ ($\bdag$)~\cite{Safavi-Naeini2013a} (see App.~\ref{sec:appF} for details). Sending the cavity output through a series of narrowband optical filters centered on the cavity resonance, as shown in Fig.~\ref{fig:setup}a, suppresses the pump so that photon counting events will correspond directly to counting phonon absorption (emission) events~\cite{Fetter1965}. Subsequently directing the filter output to a Hanbury Brown and Twiss (HBT) set-up in order to measure the second-order photon correlation function $\gTwoTau$~\cite{HanburyBrown1956b} will then result in a direct measurement of the normally (anti-normally) ordered second-order phonon correlation function.

As described in Ref.~\cite{Chan2012}, the nanobeam is patterned in such a way as to support a ``breathing" acoustic resonance at $\omegamO/2\pi = 5.6$~GHz as well as a fundamental optical resonance at a free-space wavelength near $1550$~nm, with a theoretical vacuum coupling rate of $\gzeroO/2\pi = 860$~kHz. As detailed in Ref.~\cite{Meenehan2014}, coupling to the optical resonance is accomplished via a single-sided end-fire coupling scheme using a lensed optical fiber. All device measurements presented here are peformed at room temperature and pressure.  The thermal Brownian motion of the acoustic resonance can be observed as a Lorentzian response centered around $\omegamO$ in the noise power spectral density (NPSD) $S(\omega)$ of the photocurrent of the reflected optical signal from the cavity, as shown in Fig.~\ref{fig:setup}b. The linewidth of this Lorentzian is given by $\gamma = \gammaiO + \gammaOMO$, where $\gammaiO$ is the intrinsic acoustic energy damping rate and $\gammaOMO = \pm 4 G^2 / \kappa$ is the optomechanically induced damping rate due to dynamical backaction when pumping on the red or blue sideband, respectively~\cite{Safavi-Naeini2013a}. By measuring this linewidth as a function of $\ncavO$ for both red and blue detuning, we can extract $\gammaiO/2\pi = 3$~MHz and $\gzeroO/2\pi = 645$~kHz. The optical reflection spectrum of the cavity is shown in Fig.~\ref{fig:setup}c, with the frequencies used for red and blue detuned pumping indicated by dashed lines. For this device, we observe a total optical energy decay rate of $\kappa/2\pi = 817$~MHz and a decay rate into the detection channel of $\kappae/2\pi = 425$~MHz. 

%Due to the weak nature of the optomechanical scattering process, the sideband photon emission rate is reduced relative to the pump photon flux by $50$~dB, and multiple filters must be used to suppress the pump light. In this work we use two tunable Fabry-Perot filters (Micron Optics, FFP-TF2) with bandwidths of $50$~MHz and free spectral ranges of $20$~GHz. The measured transmission spectra of each individual filter and of the two cascaded filters are shown in Fig.~\ref{fig:setup}d. At the sideband pump frequencies indicated in Fig.~\ref{fig:setup}c the total filter transmission is reduced by $81$~dB relative to the cavity resonance frequency. The single photon detectors used in this work are fiber-coupled, WSi-based superconducting nanowire single-photon detectors (SPDs) operated at $700$~mK in a helium dilution refrigerator~\cite{Marsili2013}. Each detector has a system detection efficiency $\etaSPD \approx 70\%$ at a wavelength of $1550$~nm with dark count rates of roughly $4$~Hz. Further details about the detection setup, optical and mechanical spectroscopy, and measurement of the cavity parameters can be found in the Supplementary Information (SI).

\begin{figure}[btp]
\begin{center}
\includegraphics[width=\columnwidth]{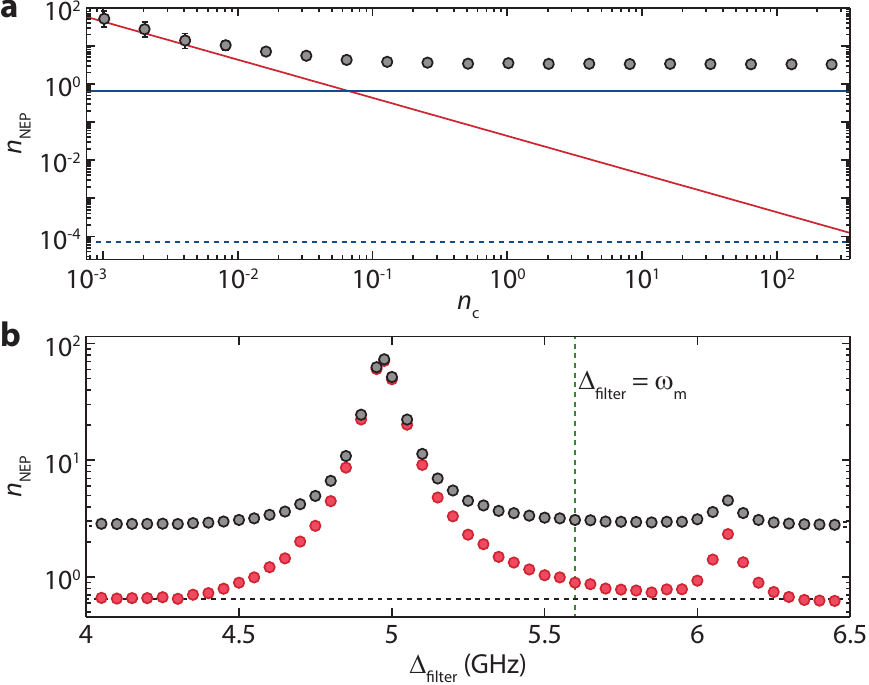}
\caption{\textbf{a}, Noise equivalent phonon number $\nNEP$ versus intracavity photon number $\ncavO$ calculated using the measured signal and noise count rates for our current setup (black circles). Solid lines indicate the theoretically expected contributions due to dark counts (red) and pump bleed-through (blue), based on the measured system efficiency and pump suppression. The blue dashed line indicates the expected pump bleed-through contribution with an additional filter. \textbf{b}, $\nNEP$ versus filter-pump detuning $\Delta_\text{filter}$ for $\ncavO \approx 65$, with (red) and without (black) an additional C-band band-pass filter inserted. The vertical green line indicates the detuning corresponding to the data from \textbf{a}, and the horizontal black line indicates the expected limiting sensitivity.} \label{fig:sensitivity}
\end{center} 
\end{figure}

%A key advantage of the phonon counting measurement presented in this work over conventional continuous measurements of position is the achievable measurement sensitivity. Typically, position is measured in optomechanical systems via heterodyne measurement of fluctuations in the optical output field. The sensitivity of these measurements is fundamentally limited by the shot noise of the probe field~\cite{Clerk2010}. In particular, for sideband resolved measurements using a red-detuned pump, it can be shown that the noise-equivalent phonon number $\nNEP$ for an ideal, shot-noise limited measurement is $\nNEP = (1+C)/4C$, where $C = |\gammaOMO|/\gammaiO$ is the cooperativity~\cite{Chan2011,Teufel2011b}. 

The sensitivity of the phonon counting scheme presented here depends upon the above-measured optomechanical cavity parameters, the dark count rate of the photon detectors, $\gammaDark$, and the residual pump laser light which is transmitted through the filters.  In this work we use a cascaded pair of tunable Fabry-Perot filters with bandwidth of $50$~MHz and free spectral range of $20$~GHz (see Fig.~\ref{fig:setup}d). The single photon detectors are fiber-coupled, WSi-based superconducting nanowire single-photon detectors (SPDs)~\cite{Marsili2013}. Each detector has a system detection efficiency $\etaSPD \approx 70\%$ at a wavelength of $1550$~nm with dark count rates of roughly $4$~Hz.  A noise-equivalent intracavity phonon occupancy ($\nNEP$) can be defined by dividing the noise count rates by the per-phonon sideband photon count rate, which yields

\begin{equation}
\nNEP = \frac{\kappa^2 \gammaDark}{4 \eta \kappae \gzeroO^2 \ncavO} + A \left( \frac{\kappa \omegamO}{2 \kappae \gzeroO}\right)^2, \label{eqn:nnep}
\end{equation} 

\noindent where $\eta$ is the total efficiency of the setup, including the system efficiency of the SPDs as well as optical insertion loss along the path from cavity to detectors, and $A$ is the transmission of the filters at the pump frequency relative to the peak transmission. The above equation makes clear the benefits of large cavity-enhanced optomechanical coupling $\gzeroO$, both in terms of the low power sensitivity limited by detector dark counts and the high power sensitivity limited by pump bleed-through.  Further details about the detection setup, optical and mechanical spectroscopy, and derivation of $\nNEP$ can be found in the App.~\ref{sec:appA}-\ref{sec:appF} below.  

In Fig.~\ref{fig:sensitivity}a we show the measured sensitivity of the phonon counting setup for $\Delta = \omegamO$ (solid grey circles), as well as the expected theoretical curve given by Eqn.~\ref{eqn:nnep}. To measure the total noise count rate, the pump beam is detuned far off-resonance from the optical cavity, $\Delta \gg \omegamO$, such that negligible motional sideband photons are generated. The total $\nNEP$ can then be computed by normalizing the measured count rate at $\Delta \gg \omegamO$ by the per-phonon sideband count rate at $\Delta = \omegamO$.  Here the per-phonon sideband count rate is extrapolated from the measured sideband count rate at low $\ncavO$, where back-action is negligible and $\nbar$ is equal to the room temperature thermal mode occupancy of $\nbath \approx 1100$. The measured sensitivity follows the expected curve at low power due to detector dark counts (solid red curve), but at high $\ncavO$ saturates to a value several times larger than expected for the filter suppression of the pump (solid blue curve).  In order to better understand this excess noise, Fig.~\ref{fig:sensitivity}b shows measurements of the $\nNEP$ as a function of filter-pump detuning, $\Delta_\text{filter}$, at a high power where the pump transmission dominates the noise ($\ncavO \approx 65$). A strong dependence on $\Delta_\text{filter}$ is observed, with a peak in the noise at $5$~GHz and a secondary peak at $6.1$~GHz, consistent with laser phase-noise of our pump laser~\cite{Safavi-Naeini2013a}.  With the addition of a C-band bandpass filter prior to the SPD to remove broadband spontaneous emission from the pump laser, and at frequencies far from the laser phase-noise peaks, the measured $\nNEP$ agrees well with the theoretical predictions based on the filter pump suppression (horizontal dashed curve).  At the relevant detuning of $\Delta_\text{filter} = \omegamO$ (vertical dashed curve), we measure a limiting sensitivity of $\nNEP = 0.89 \pm 0.05$.

%. The remaining laser noise can be removed in future experiments by filtering the pump laser prior to the cavity, but does not limit our measurements in this work due to the large thermal phonon occupancy ($\nbar \sim 1100$) at room temperature. 

%While our current setup achieves sensitivities below the single phonon level at high power, by removing this excess noise and adding a third post-cavity filter, the sensitivity of this detection scheme can be pushed well below the shot-noise limit even for modest intracavity photon numbers $\ncavO \sim 1$.

\begin{figure}[btp]
\begin{center}
\includegraphics[width=\columnwidth]{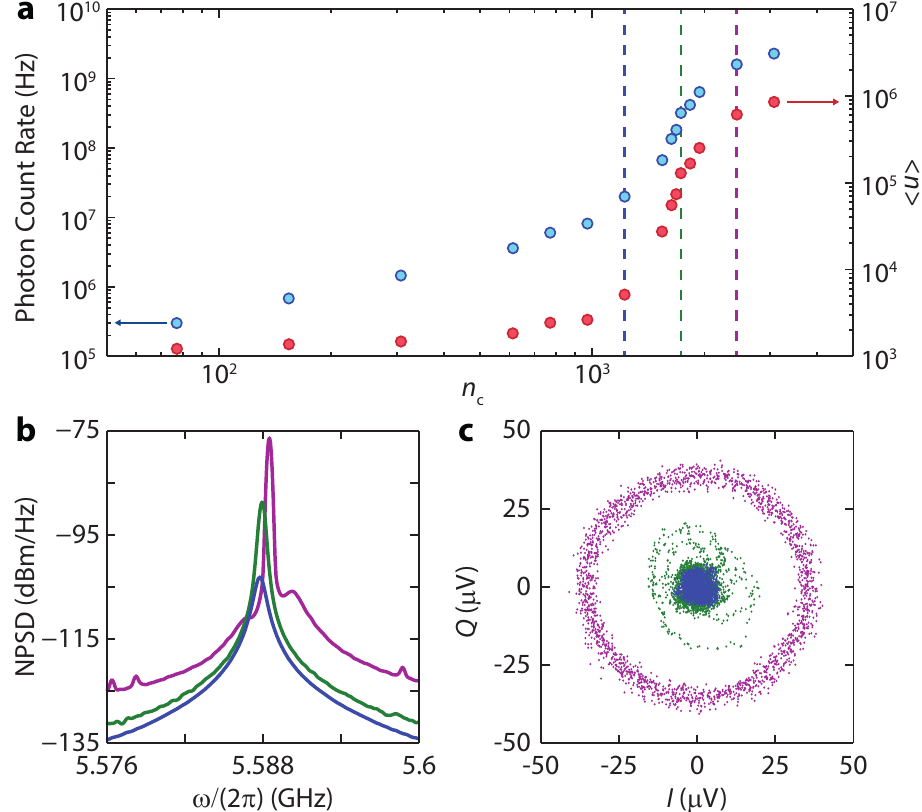}
\caption{\textbf{a}, Phonon count rate (blue) and inferred phonon occupancy $\nbar$ (red) as a function of intracavity photon number for $\Delta = -\omegamO$. Dashed lines indicate points below (blue), at (green), and above (magenta) threshold. \textbf{b}, Noise power spectral densities (NPSD) corresponding to the dashed lines in \textbf{a}. \textbf{c}, Phase plots of the in-phase ($I$) and in-quadrature ($Q$) amplitudes of the detected optical noise corresponding to the dashed lines in \textbf{a}} \label{fig:count_rate}
\end{center} 
\end{figure}

In what follows, we focus on measurements made with a blue-detuned pump ($\Delta = -\omegamO$), in which the optomechanical back-action results in instability and self-oscillation of the acoustic resonator~\cite{Grudinin2010,Safavi-Naeini2013a}. The Stokes sideband count rate detected on a single SPD, shown versus $\ncavO$ in Fig.~\ref{fig:count_rate}a, displays a pronounced threshold, with an exponential increase in output power beginning at $\ncavO \approx 1200$, where $C\equiv |\gammaOMO|/\gammaiO \approx 0.8$, in agreement with the expected onset of instability around $C = 1$ ($\gamma = 0$). This sharp oscillation threshold can also be observed from the measured NPSD (Fig.~\ref{fig:count_rate}b), in which the amplitude of the mechanical spectrum is seen to rapidly increase with a simultaneous reduction in linewidth, and in plots of the in-phase and in-quadrature components of the photocurrent fluctuations, which show a transition from thermal noise to a large amplitude sinusoidal oscillation. Also shown in Fig.~\ref{fig:count_rate}a is the inferred phonon occupancy $\nbar$, determined by calculating the per-phonon sideband scattering rate at low $\ncavO$, where $\nbar$ is roughly equal to the thermal occupation, and extrapolating to higher powers assuming linearity in the optomechanical interaction.  As detailed in App.~\ref{sec:appG}, a self-consistent determination of the oscillation amplitude, based upon comparison of the sideband count rates above and below threshold, indicates that even at our highest pump power the mechanical amplitude remains small enough that the linear approximation remains valid.

%Above threshold, the amplitude of self-oscillation can settle into one of a number of stable attractors~\cite{Marquardt2006}. In principle this amplitude can become large enough that the optomechanical interaction can no longer be approximated as linear, in which case there is no longer a one-to-one correspondence between the creation of sideband photons and phonons. 

\begin{figure}[btp]
\begin{center}
\includegraphics[width=\columnwidth]{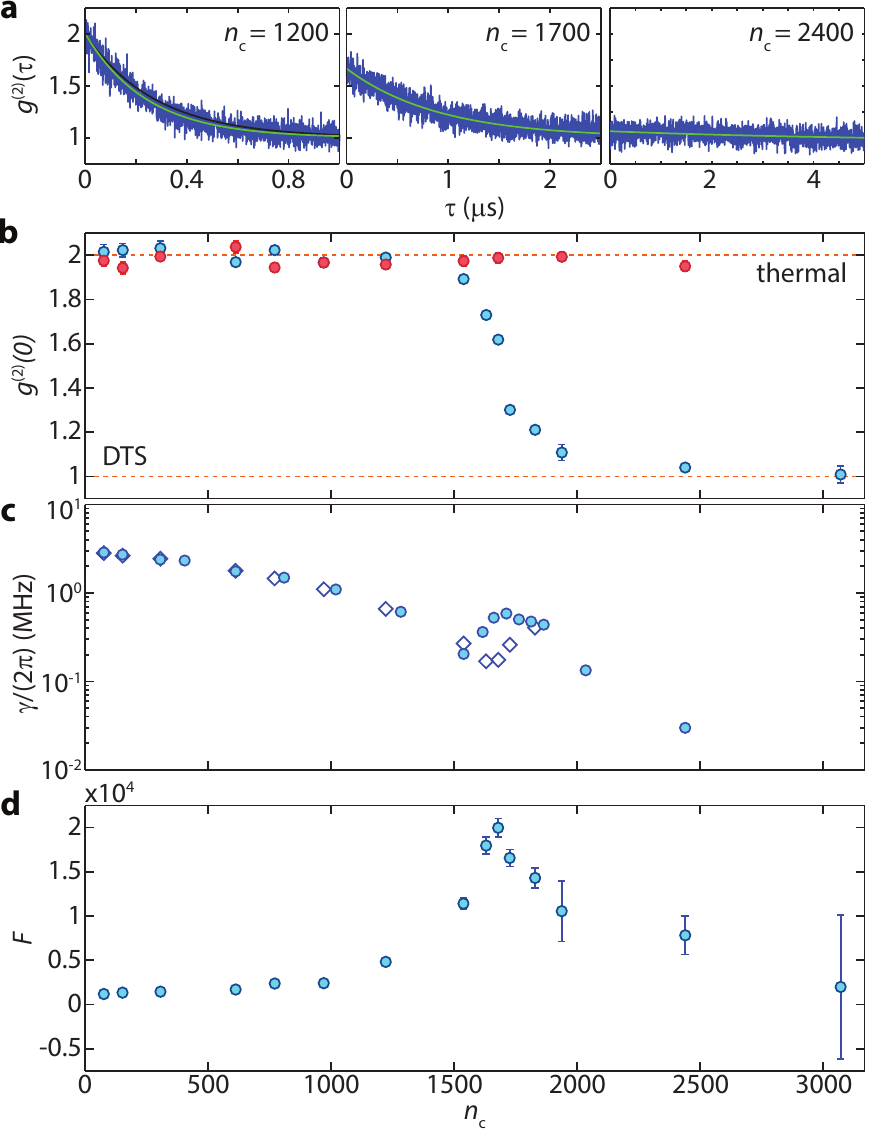}
\caption{\textbf{a}, Normalized second-order intensity correlation function $\gTwoTau$ for $\Delta = -\omegamO$, shown below, at and above threshold (left to right, respectively). Green lines show a simple exponential fit, while black lines indicate the expected theoretical curve using decay rates measured from fitting the NPSD linewidth. \textbf{b}, Phonon correlation at zero time delay versus $\ncavO$ for $\Delta = -\omegamO$ (blue) and $\Delta = \omegamO$ (red). The top and bottom dashed lines indicate the expected values for purely thermal or displaced thermal states (DTS), respectively. \textbf{c}, Mechanical decay rate versus $\ncavO$ for $\Delta = -\omegamO$, determined from the measured linewidth of the NPSD (circles) and from the exponential fit to $\gTwoTau$ (diamonds). \textbf{d}, Fano factor versus $\ncavO$.} \label{fig:g2}
\end{center} 
\end{figure}

The statistical properties of the resonator near the self-oscillation threshold can also be characterized by measuring photon correlations using a HBT setup as shown in Fig.~\ref{fig:setup}. As the oscillation threshold is crossed, the state of the acoustic resonator will transition from a thermal state into a displaced thermal state (DTS), and the normalized phonon intensity correlation function near $\tau = 0$ should show a transition from bunching ($\gTwoZero > 1$) to Poissonian statistics ($\gTwoTau = 1$ for all $\tau$). Plots of $\gTwoTau$ below, at, and above threshold are shown in Fig.~\ref{fig:g2}a. Below threshold bunching is clearly visible, with $\gTwoZero = 2$ as expected for a purely thermal state.  In Fig.~\ref{fig:g2}b $\gTwoZero$ is plotted versus $\ncavO$ for both blue- and red-detuned pump light. For a blue-detuned pump a smooth decrease from $\gTwoZero = 2$ to $\gTwoZero = 1$ is observed in the threshold region, while for a red-detuned pump the oscillator is observed to remain in a thermal state through threshold and beyond. The decay rate of the acoustic resonator, measured from both the linewidth of the NPSD and from an exponential fit to $\gTwoTau$ below threshold, is plotted in Fig.~\ref{fig:g2}c. The decay rate as measured from the NPSD, which includes both phase and amplitude fluctuations, is seen to increase around threshold before continuing to decrease. This behavior is commonly observed in semiconductor lasers where a coupling exists between the gain and the cavity refractive index, and a similar effect arises in optomechanical oscillators due to the optical spring effect~\cite{Rodrigues2010}. The decay rate measured from $\gTwoTau$, on the other hand, which measures intensity fluctuations, begins to deviate from the measured linewidth in the vicinity of threshold.  Thermal phonon emission dictates a strict correspondence between the second-order and first-order coherence functions~\cite{Glauber1963c}, however, above threshold where the phonon statistics are no longer purely thermal, such a deviation is possible, and in fact predicted for self-sustaining oscillators~\cite{Lax1967c}. The Fano factor, defined as $F = (\Delta n)^2/\nbar = 1+\nbar (\gTwoZero -1)$, provides additional statistical information about the fluctuations of the oscillator, and is useful for defining a precise oscillator threshold~\cite{Rice1994} as well as distinguishing between states that may have similar or identical values of $\gTwoZero$ (e.g. a coherent state versus a DTS)~\cite{Rodrigues2010}. The Fano factor of our mechanical oscillator, computed from the measured $\gTwoZero$ and the inferred values of $\nbar$, is displayed in Fig.~\ref{fig:g2}d and shows the expected increase and peak in fluctuations at threshold. Above threshold, the Fano factor drops again due to saturation in the optomechanical gain, approaching a measured value consistent with that expected for a DTS ($F \sim 2\nbath + 1$).

%It should be noted that this provides unambiguous proof that this technique measures the phonon correlation function directly, as the count rates are several orders of magnitude larger than the dark count rates and measurement of the pump bleed-through would not display bunching.

Although we have emphasized the analogy between the optomechanical oscillator studied here and a laser, there are unique differences which arise due to the intrinsically nonlinear nature of the radiation pressure interaction in an optomechanical cavity.  Recent theoretical studies~\cite{Rodrigues2010,Qian2012,Loerch2014} indicate that a laser-driven optomechanical oscillator will enter a nonclassical mechanical state with anti-bunched phonon statistics ($F < 1$), and under slightly more restrictive conditions, strongly negative Wigner density.  Surprisingly, this is predicted to be observable even for classical parameters, i.e., outside the single-photon strong-coupling regime ($\gzeroO/\kappa  < 1$), and in the presence of thermal noise.  Beyond phonon correlation spectroscopy of optomechanical oscillators, it is envisioned that sensitive photon counting of the filtered motional sidebands may be utilized in the preparation and heralding of non-Gaussian quantum states of a mechanical resonator~\cite{Vanner2013}.  For the OMC cavities of this work, with their large optomechanical coupling rate and near millisecond-long thermal decoherence time at sub-Kelvin temperatures~\cite{Meenehan2014}, the phonon addition and subtraction processes of Ref.~\cite{Vanner2013} should be realizable with high fidelity and at rates approaching a megahertz.  Whether for studies of the quantum behavior of mesoscopic mechanical objects or in the context of proposed quantum information processing architectures utilizing phonons and photons~\cite{Stannigel2012}, such photon counting methods are an attractive way of introducing a quantum nonlinearity into the cavity-optomechanical system.      

%it is envisioned that sensitive photon/phonon counting techniques will also be valuable in studying the temporal dynamics of optomechanical resonators interacting with their environment, whether it be to explore the physics of phonon damping in such structures~\cite{Meenehan2014}, or simply as a high-bandwidth sensor~\cite{}. 

\begin{acknowledgements}
The authors would like to thank F. Marquardt and A. G. Krause for helpful discussions, as well as V. B. Verma, R. P. Miriam and S. W. Nam for their support with the single photon detectors used in this work. This work was supported by the DARPA ORCHID and MESO programs, the Institute for Quantum Information and Matter, an NSF Physics Frontiers Center with support of the Gordon and Betty Moore Foundation, and the Kavli Nanoscience Institute at Caltech. Part of this work was carried out at the Jet Propulsion Laboratory, under contract with the National Aeronautics and Space Administration.  ASN acknowledges support from NSERC.  SG was supported by a Marie Curie International Outgoing Fellowship within the $7^{th}$ European Community Framework Programme.
\end{acknowledgements}

%\bibliography{../Mirror_v2_OJP}

\appendix

\section{Optical and mechanical design}
\label{sec:appA}

\begin{figure}[]
\begin{center}
\includegraphics[width=\columnwidth]{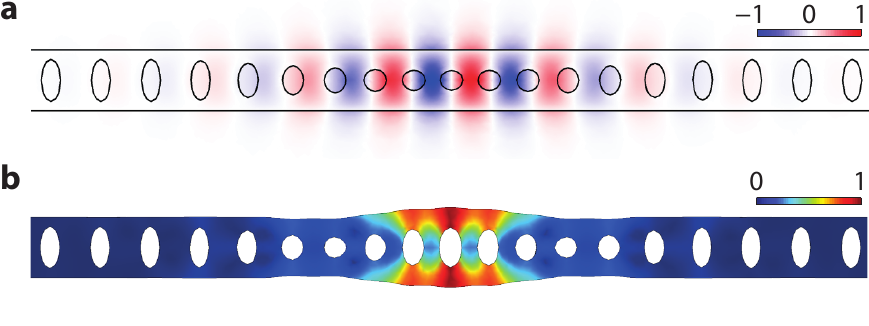}
\caption{\textbf{FEM simulations.} \textbf{a}, Electric field $E_{\text{y}}$ component of the optical mode at frequency $\omegacO/2\pi = 194$~THz (polarization in the plane of the page and transverse to the long axis of the nanobeam). \textbf{b} Displacement field of the mechanical breathing mode at frequency $\omegamO/2\pi = 5.1$~GHz.} \label{Sfig:modes}
\end{center} 
\end{figure}

The optomechanical crystal (OMC) studied in this work is numerically optimized for both optomechanical coupling and optical/acoustic quality factor, via finite-element method (FEM) simulation in COMSOL Multiphysics~\cite{COMSOL}, according to the procedure outlined in Ref.~\cite{Chan2012}. The holes on the ends of the beam support simultaneous bandgaps for optical light with wavelengths near $1550$~nm and acoustic waves with frequencies from $3-5.5$~GHz, while the variation of holes towards the center of the beam perturb the bandgaps so as to create co-localized optical and acoustic midgap resonances. The fundamental optical mode (Fig.~\ref{Sfig:modes}a) has a nominal wavelength of $1545$~nm and the fundamental acoustic mode (Fig.~\ref{Sfig:modes}b) has a nominal resonance frequency of $5.1$~GHz. The nominal optomechanical vacuum coupling rate, due predominantly to photoelastic effects, is predicted to be $\gzeroO/2\pi = 860$~kHz. Physically, this coupling rate is the optical resonance frequency shift due to the zero-point fluctuations of the acoustic resonator. 

\section{Fabrication}
\label{sec:appB}

The devices are fabricated from a silicon-on-insulator (SOI) wafer (SOITEC, $220$~nm device layer, $3 \mu$m buried oxide) using electron beam lithography followed by reactive ion etching (RIE/ICP). The Si device layer is then masked using a ProTEK PSB photoresist to define a mesa region of the chip to which a tapered lensed fiber can access. Outside of the protected mesa region, the buried oxide is removed with a plasma etch and a trench is formed in the underlying silicon substrate using tetramethylammonium hydroxide (TMAH). The devices are then released in hydrofluoric acid ($49\%$ aqueous HF solution) and cleaned in a piranha solution ($3$-to-$1$ H$_2$SO$_4$:H$_2$O$_2$) before a final hydrogen termination in diluted HF. In fabrication, arrays of the nominal design shown in Fig.~\ref{Sfig:modes} are scaled by $\pm2\%$ to account for frequency shifts due to fabrication imperfections and disorder.

\section{Experimental setup}
\label{sec:appC}

\begin{figure}[btp]
\begin{center}
\includegraphics[width=\columnwidth]{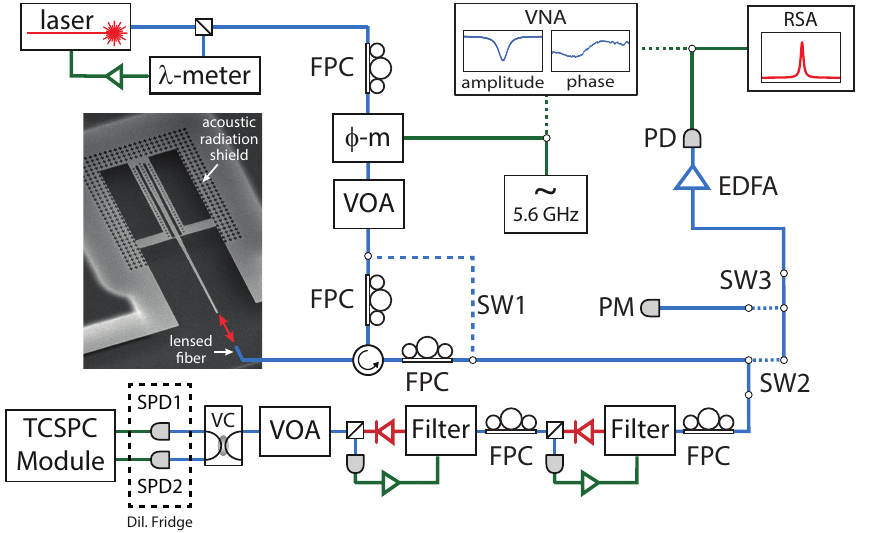}
\caption{\textbf{Experimental setup.} Phonon counting setup. $\lambda$-meter: wavemeter, FPC: fiber polarization controller, $\phi$-m: electro-optic phase modulator, VOA: variable optical attenuator, SW: optical switch, PM: optical power meter, EDFA: erbium-doped fiber amplifier,  PD: fast photodiode, RSA: real-time spectrum analyzer, VNA: vector network analyzer, VC: variable coupler, SPD: superconducting single photon detector, TCSPC: time-correlated single photon counting.} \label{Sfig:setup}
\end{center} 
\end{figure}

The full experimental setup for phonon counting and intensity interferometry is shown in Fig.~\ref{Sfig:setup}. A fiber-coupled, wavelength tunable external cavity diode laser is used as the light source, with a small portion of the laser output sent to a wavemeter ($\lambda$-meter) for frequency stabilization. The remaining laser power is sent through an electro-optic phase modulator ($\phi$-m), used to generate optical sidebands for locking the filter cavities, and a variable optical attenuator (VOA) to allow control of the input power to the cavity. The signal is then sent into an optical circulator which sends the optical probe to the a lensed fiber tip for end-fire coupling to the device. The cavity reflection can then be switched to one of two detection setups. The first allows the signal to be switched to a power meter (PM) for measuring the reflected signal power or to an erbium-doped fiber amplifier (EDFA) followed by a high-speed photodetector (PD). The resulting photocurrent can be sent to a real-time spectrum analyzer (RSA) in order to measure the noise power spectral density (NPSD) of the optical signal or to a vector network analyzer (VNA) which can be used to measure the full complex response of the optical cavity. The second detection setup sends the cavity reflection through a series of narrowband tunable Fabry-Perot filters ($\sim50$~MHz bandwidth, $\sim20$~GHz free-spectral range) in order to reject the pump frequency. The signal then travels through a variable coupler (VC) and is sent to the dilution refrigerator where is is detected by two superconducting single photon detectors (SPD). The output of these detectors is sent to a time-correlated single photon counting (TCSPC) module for calculation of the detection correlation function. 

Since the pump laser is tuned to a motional sideband during the phonon counting measurement, the two Fabry-Perot filters used in this work must be tuned to the optical cavity resonance via an initial lock and stabilization procedure. This procedure is as follows.  First, the pump wavelength is tuned to the blue or red OMC sideband by optimizing the mechanical transduction signal on a spectrum analyzer. Since the power of the radiated Stokes- or anti-Stokes-scattered light is too low to provide a feedback signal for filter stabilization, we then bypass the OMC and phase-modulate the pump at frequency $\omegamO$. The filters are then locked to maximize transmission of the sideband which is resonant with the cavity. After a stabilization period of a few seconds, the filter positions are held without further feedback while the pump modulation is turned off, pump power is adjusted for the desired $\ncavO$, and the OMC is switched back into the optical path. Once locked, the transmission of the filters is observed to be stable to within $5-10\%$ for several minutes in the absence of active feedback locking.

In order to avoid pile-up artifacts~\cite{Becker2005} in the acquired $\gTwoTau$ histograms presented in the main text, the photon count rate incident upon the SPDs is kept at or below $30$~kHz. This is accomplished with a variable optical attenuator on the output of the filters, and is sufficient to maintain a flat histogram background over a $5$~$\mu$s window. The absolute count rate reported in Fig.~\ref{fig:count_rate}a takes the variable attenuation into account.

\section{Device characterization}
\label{sec:appD}

\begin{figure}[btp]
\begin{center}
\includegraphics[width=\columnwidth]{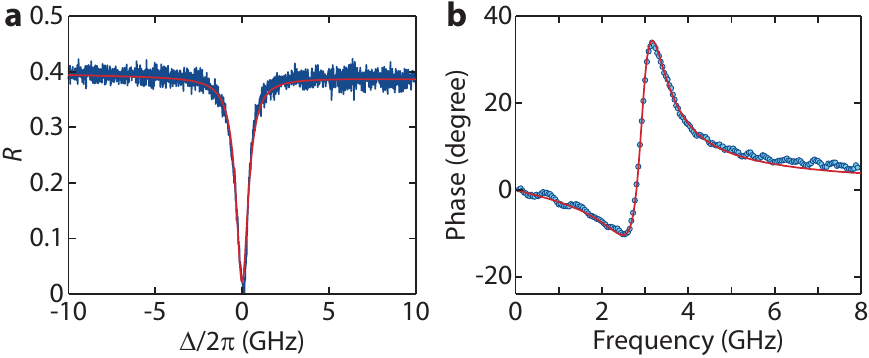}
\caption{\textbf{Optical cavity response.} \textbf{a}, Optical reflection spectrum of the cavity resonance versus cavity detuning $\Delta = \omegacO-\omegalO$ (blue) with a Lorentzian fit (red) yielding $\QoO = 236,000$. \textbf{b}, Phase response of the optical resonance, yielding $\kappae/\kappa = 0.52$.} \label{Sfig:optical_scan}
\end{center}
\end{figure}

Full characterization of the optical resonance involves determination of the single pass fiber-to-waveguide coupling efficiency $\eta_{\text{cpl}}$, the total energy decay rate $\kappa$, and the waveguide-cavity coupling efficiency $\eta_\kappa = \kappae/\kappa$ ($\kappae$ is the decay rate into the  detection channel). The fiber collection efficiency is determined by observing the calibrated reflection level far-off resonance with the cavity and is found to be $\eta_\text{cpl} = 0.63$. The total cavity decay rate is determined by fitting the optical reflection spectrum of the cavity (Fig.~\ref{Sfig:optical_scan}a) and yields $\kappa/2\pi = 818$~MHz (optical quality factor $\QoO = 236,000$). The reflection level on resonance, when normalized to the off-resonance reflection level, is related to the cavity-waveguide coupling efficiency by $R_0 = (1-2\eta_\kappa)^2$. However, for single-sided coupling this is not a single-valued function of the coupling efficiency. Consequently, the complex response of the cavity is measured by locking the laser off-resonance from the cavity and using a vector network analyzer (VNA) to drive an electro-optic modulator (EOM) and sweep an optical sideband across the cavity. By detecting the reflected power on a high-speed photodetector connected to the VNA input, the phase response of the cavity can be measured (Fig.~\ref{Sfig:optical_scan}b). Fitting this with prior knowledge of the cavity resonance frequency and decay rate yields $\eta_\kappa = 0.52$.

\begin{figure}[btp]
\begin{center}
\includegraphics[width=\columnwidth]{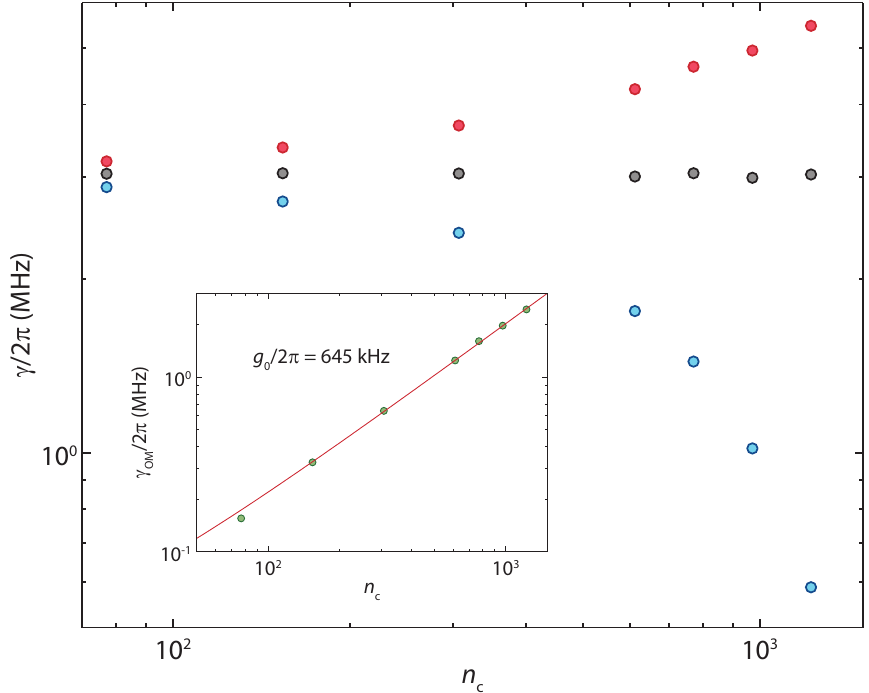}
\caption{\textbf{Calibration of $\gzeroO$.} Mechanical linewidth $\gamma$ versus intracavity photon number $\ncavO$ for $\Delta = \omegamO$ (red) and $\Delta = -\omegamO$ (blue). The intrinsic linewidth of the acoustic resonator $\gammaiO$ (black) is determined by averaging the blue detuned data and yields $\QmO = 1850$. The inset shows the optomechanically induced damping $\gammaOMO$, obtained by subtracting $\gammaiO$ from $\gamma$, versus $\ncavO$. The linear fit shown in red yields a vacuum optomechanical coupling rate of $\gzeroO = 645$~kHz.}\label{Sfig:gammaRSA}
\end{center}
\end{figure}

To characterize the acoustic resonance, the cavity reflection is sent through an erbium-doped fiber amplifier (EDFA) and detected on a high-speed photodetector. The EDFA is used to amplify the signal so that the optical noise floor overcomes the detector's electronic noise, and the noise power spectral density (NPSD) of the optical reflection is measured on a real-time spectrum analyzer (RSA), where a Lorentzian response due to transduction of the acoustic thermal Brownian motion can be observed at the acoustic resonant frequency $\omegamO/2\pi = 5.6$~GHz. For a pump laser locked onto the red or blue mechanical sideband of the cavity ($\Delta = \omegacO - \omegalO = \pm \omegamO$) the linewidth of the transduced signal is given by $\gamma = \gammaiO \pm \gammaOMO$, where $\gammaOMO = \pm 4 \gzeroO^2 \ncavO / \kappa$ ($\ncavO$ is the steady state intracavity photon number). The dependence of linewidth for both detunings versus $\ncavO$ is shown in Fig.~\ref{Sfig:gammaRSA}. By averaging the two sets of data we can extract the intrinsic acoustic damping rate $\gammaiO/2\pi = 3$~MHz ($\QmO = 1850$), which is seen to remain constant as a function of $\ncavO$. By fitting the excess optomechanically induced damping $\gammaOMO$ as a function of $\ncavO$ we extract a coupling rate of $\gzeroO = 645$~kHz.

\section{Single photon detectors}
\label{sec:appE}

The detectors used in this work are amorphous WSi-based superconducting nanowire single-photon detectors (SNSPDs, or SPDs hereafter) developed in collaboration between the Jet Propulsion Laboratory and NIST. The SPDs are designed for high-efficiency detection of individual photons in the wavelength range $\lambda = 1520 - 1610$~nm with maximum count rates of about $25 \times 10^{6}$~counts per second (c.p.s). (reset time $t_R = 40$~ns)~\cite{Marsili2013} and very low dark count rates (DCRs). With the SPD in the superconducting state (below its critical temperature $\Tc = 3.7$~K), a DC bias current $\Ib$ of a few microamps is maintained through the nanowire by an external current source. The operating range of the detector lies between a lower cutoff current and an upper switching current $\Isw$, above which the nanowire switches to a non-superconducting state. The choice of quiescent operating current $\Ib$ for each detector was made to roughly maximize the ratio of the quantum efficiency $\etaSPD$ to DCR while operating within the "plateau" region where both $\etaSPD$ and the DCR are nearly constant as $\Ib$ is adjusted slightly.

The SPDs are mounted on the still stage of a $^{3}$He/$^{4}$He dilution refrigerator at 700 mK. Single-mode optical fibers (Corning SMF-28) are passed into the refrigerator through vacuum feed-throughs and coupled to the SPDs via a fiber sleeve attached to each SPD mount. Proper alignment of the incoming fiber with the $15$ $\mu$m $\times$ $15$ $\mu$m square area of the SPD nanowire is ensured by a self-aligned mounting system incorporated into the design of the SPD~\cite{Marsili2013}. The radio-frequency output of each SPD is amplified by a cold-amplifier mounted on the $50$~K stage of the refrigerator as well as a room-temperature amplifier, then read out by a triggered PicoQuant PicoHarp 300 time-correlated single photon counting module. The counting module is triggered by input pulses reaching a voltage above a fixed discriminator value $\Vd$. Amplified photon-detection pulse heights of $150 - 250$~mV are typical, and corresponding discriminator values in the range $110 - 150$~mV were chosen for each SPD by measuring nominal count rates as a function of $\Vd$ and choosing an operating value of $\Vd$ to be near the center of the plateau region in which the observed count rates are independent of small changes in the discriminator setting.

Initial characterization of the SPDs was centered on measuring dark count rates and the quantum efficiencies of the detectors. The measured DCRs are sensitive to various channels by which stray light may couple into the fiber-detector system, including ambient laboratory lighting and thermal radiation both inside and outside the refrigerator. By tightly spooling ($\sim1.5$~inch diameter) the optical fiber within the fridge to filter out long wavelength blackbody radiation and systematically isolating the optical fiber from environmental light sources we have achieved DCRs of $2-4$~c.p.s.

Quantum efficiency measurements of the SPDs were made using laser light of $\lambda = 1554$~nm attenuated to an input power of $1.53$~fW at the input to the fridge, corresponding to $N \approx 12,000$ incoming photons per second. We calculate $\etaSPD$ by referring the detected photon count rate (less the corresponding known DCR) to the nominal input flux $N$. This efficiency incorporates the intrinsic detection efficiency of the SPDs as well as any losses in the fiber run within the fridge and in the coupling between the fiber and the SPD itself. At just below the respective switching currents of the detectors, we find $\etaSPD = 70\%$, with this result depending on photon polarization ($\lesssim 20\%$ variability).

\section{Phonon counting sensitivity}
\label{sec:appF}
For sufficiently weak optomechanical coupling ($\gzeroO \ll \kappa$) and small mechanical amplitude, the equations of motion for the optomechanical system can be linearized about a large steady state optical field amplitude. For a sideband resolved system ($\kappa/2 \ll \omegamO$) and a red-detuned pump ($\Delta \approx \omegamO$) the output optical field may then be written in the Fourier domain (in a frame rotating at the pump frequency) as~\cite{Safavi-Naeini2013a}

\begin{eqnarray}
\hat{a}_\text{out}(\omega) & = & \left(1-\frac{\kappae}{i(\Delta-\omega)+\kappa/2} \right) \hat{a}_\text{in}(\omega) \notag \\ &&-\frac{\sqrt{\kappae\kappai}}{i(\Delta-\omega)+\kappa/2} \hat{a}_\text{i}(\omega) \notag \\
&&-i\frac{\sqrt{\kappae \ncavO}\gzeroO}{i(\Delta-\omega)+\kappa/2} \hat{b}(\omega),
\end{eqnarray}

\noindent where $\hat{a}_\text{in}(\omega) = \alpha \delta(\omega) + \hat{a}_\text{vac}(\omega)$ ($\alpha$ is the steady-state optical field at the pump frequency, $\hat{a}_\text{vac}(\omega)$ is the vacuum noise of the pump), $\kappai = \kappa-\kappae$ is the intrinsic loss rate of the optical cavity, $\hat{a}_\text{i}(\omega)$ is additional vacuum noise admitted via the intrinsic loss channels, and $\hat{b}(\omega)$ is the annihilation operator for the acoustic resonator. Note that $\hat{b}^{\dagger}(\omega)$ takes the place of $\hat{b}(\omega)$ for a blue-detuned pump ($\Delta \approx -\omegamO$).

As $\hat{b}(\omega)$ is sharply peaked around $\omega = \omegamO$, we can spectrally filter out the strong optical pump at $\omega = 0$. The additional optical noise, assumed to be white Gaussian noise, cannot be filtered out in this way. However, in the case that the optical noise is pure vacuum noise it will not contribute to any photon counting events. Thus, for the purposes of photon counting the output optical field can be written post-filtering as

\begin{equation}
\hat{a}_\text{out}(t) \approx \frac{2 \sqrt{\kappae \ncavO} \gzeroO}{\kappa} \hat{b}(t) = \sqrt{\frac{\kappae}{\kappa}} \sqrt{\gammaOMO} \hat{b}(t),
\end{equation}

\noindent which shows explicitly that in this linearized regime photon counting is equivalent to phonon counting ($\langle \hat{a}^{\dagger}_\text{out} \hat{a}_\text{out} \rangle \propto \langle \hat{b}^{\dagger} \hat{b} \rangle$).

As can be seen in the above equation, the optically induced acoustic damping rate $\gammaOMO$ physically represents the per-phonon rate of sideband photon emission, corresponding to phonon absorption (emission) for $\Delta = \omegamO$ ($\Delta = -\omegamO$). Of the sideband photons emitted into the optical cavity, a fraction $\kappae/\kappa$ are subsequently emitted into the detection channel and detected with overall system efficiency $\eta$, including both the system efficiency of the SPDs as well as insertion loss along the path from cavity to detector. The count rate per phonon is thus given by $\gammaSB= \eta (\kappae/\kappa) \gammaOMO$, and the total count rate is given by

\begin{equation}
\Gamma_\text{tot} = \gammaSB \nbar + \gammaPump + \gammaDark,
\end{equation}

\noindent where $\nbar$ is the average phonon occupancy of the acoustic resonator, $\gammaPump$ is the count rate due to residual pump transmission through the filters, and $\gammaDark$ is the intrinsic dark count rate of the SPD.

To assess the sensitivity of the phonon counting measurement, the noise count rate can be divided by the per-phonon sideband count rate to obtain a noise-equivalent phonon number

\begin{equation}
\nNEP = \nPump + \nDark = \frac{\gammaPump}{\gammaSB} + \frac{\gammaDark}{\gammaSB}.
\end{equation}

The dark count rate $\gammaDark$ is simply a measured constant, while $\gammaPump = \eta A \dot{N}_\text{pump}$, where $A$ is the transmission of the filters at the pump frequency relative to the peak transmission, and $\dot{N}_\text{pump}$ is the input photon flux of the pump, which is nearly perfectly reflected from the cavity when the pump is far off-resonance. For a pump detuning from the cavity of $\Delta = \omegamO$, the input photon flux can be related to the intracavity photon number $\ncavO$ by $\dot{N}_\text{pump} \approx \omegamO^2 \ncavO / \kappae$. Thus, we can write the total noise-equivalent phonon number as

\begin{equation}
\nNEP = \frac{\kappa^2 \gammaDark}{4 \eta \kappae \gzeroO^2 \ncavO} + A \left( \frac{\kappa \omegamO}{2 \kappae \gzeroO}\right)^2.
\end{equation}

\section{Oscillation amplitude above threshold}
\label{sec:appG}
The classical equation of motion for the optical cavity field $\alpha$ is given in the frame rotating at the pump frequency by

\begin{equation}
\dot{\alpha} = -\left(i \left(\Delta+\gzeroO x\right) +\frac{\kappa}{2}\right) \alpha + \Omega,
\end{equation}

\noindent where $x$ is the position of the acoustic resonator, $\Omega = \sqrt{\kappae P_\text{in} / \hbar \omegacO}$, and $P_\text{in}$ is the optical input power at the device.  In the regime of parametrically driven self-oscillation, the amplitude of the acoustic oscillator can become large enough that the usual linearization approximation becomes invalid. However, using the ansatz that the mechanical oscillation amplitude is given by $x(t) = \beta \text{sin}(\omega_m t)$ (note that $\beta$ is given in units of the zero-point amplitude, so that $\beta^2 = 4 \nbar+2$), an exact solution for the optical cavity field can be written as a sum of sidebands.

In particular, the equation of motion can be formally integrated to yield

\begin{equation}
\alpha(t) = \Omega e^{-i z \text{cos}(\omegamO t)} \int_{0}^{\infty} d\tau e^{-\left(i \Delta + \frac{\kappa}{2}\right) \tau} e^{i z \text{cos}(\omegamO (t-\tau))},
\end{equation}

\noindent where $z = \gzeroO \beta/\omegamO$. This can be solved exactly by using the Jacobi-Anger expansion $e^{i z \text{cos}(\theta)} = \sum_{n} i^n J_\text{n} (z) e^{i n \theta}$, where $J_\text{n}$ is a Bessel function of the first kind, to yield

\begin{align}
\alpha(t) & =   e^{-i z \text{cos}(\omegamO t)} \sum_n e^{i n \omegamO t} \frac{i^n \Omega J_\text{n} (z)}{\kappa/2+i(\Delta+n\omegamO)}  \notag \\
& = \sum_n \sum_m e^{i(n-m)\omegamO t} \frac{i^{n-m} \Omega J_\text{n} (z) J_\text{m} (z)}{\kappa/2+i(\Delta+n\omegamO)}, \label{eqn:sideband_deriv}
\end{align}

\noindent which can be reindexed to have the form $\alpha(t) = \sum_n \alpha_n e^{i n \omegamO t}$, with

\begin{equation}
\alpha_\text{n} = i^n \Omega \sum_m \frac{J_\text{m} (z) J_\text{m-n} (z)}{h_\text{m}}, \label{eqn:sideband_expr}
\end{equation}

\noindent where $h_\text{m} = \kappa/2 + i(\Delta + m\omegamO)$. Note that if we are only interested in the total energy in the cavity, $|\alpha(t)|^2$, the final expansion of the global phase factor in Eqn.~\ref{eqn:sideband_deriv} is unnecessary and a simpler form for $\alpha_\text{n}$ involving no sums may be used~\cite{Marquardt2006,Rodrigues2010}. However, for the case considered here where a specific frequency component of the cavity output is filtered before detection it is necessary to use the full expression given in Eqn.~\ref{eqn:sideband_expr}.  In the regime of self-oscillation, the oscillation amplitude $\beta$ is determined by balancing the optically induced mechanical gain with the intrinsic mechanical loss

\begin{equation} 
\gammaOMO = \frac{4 \gzeroO^2 \Omega^2}{\omegamO} \text{Im} \left[ \sum_\text{n} \frac{J_\text{n}(z) J_\text{n+1}(z)}{z h_\text{n} h_\text{n+1}^{*}} \right] = -\gammaiO. \label{eqn:gain_balance}
\end{equation}

The presence of the filters at the cavity output ensures that the SPD count rate is proportional to the number of intracavity photons in the first Stokes sideband at $\omega = \omegalO-\omegamO$, which is given by $n_1 = |\alpha_1|^2$. For the lowest input power, below threshold, $n_1$ is given by the simple linear approximation $n_1 = [(4G^2/\kappa)\nbath]/\kappa$, where $\nbath \approx 1100$ is the thermal occupancy of the mechanical resonator. Near and above threshold, $n_1$ can be determined from the ratio of the above and below threshold SPD count rates and the known value of $n_1$ below threshold. For a given input power with $n_1$ constrained, Eqs.~\ref{eqn:sideband_expr} and \ref{eqn:gain_balance}, along with the condition that $|\gammaOMO| \approx \gammaiO$ above threshold, can be used to solve for the detuning $\Delta$ and amplitude $\beta$ commensurate with the observed count rate.

For our highest input power, we find $\Delta \approx -1.067 \omegamO$ and $z \approx 0.15$. This amplitude is small enough that the linear approximation ($\alpha_1 \propto z$) is still valid. In particular, in the linear regime the relation $J_1(z) \approx z/2$ should hold. For the largest value of $z$ in our measurements, we find that $J_1(z)$ differs from $z/2$ by only about $0.3\%$. The shift in detuning, and the concomitant reduction in oscillation amplitude, is expected due to the thermo-optic effect which will tend to shift the cavity resonance to lower frequencies as the total intracavity photon population is increased by the amplified Stokes scattering~\cite{Krause2014}.

In the linearized regime, $\ncavO$ (the intra-cavity photon number of the 0th sideband at $\omegalO$) is to a good approximation equal to the intra-cavity photon number in the absence of optomechanical coupling, $\ncavO = (4\kappae P_{\text{in}}/\hbar\omegalO)/(\kappa^2 + 4\Delta^2)$.  The total optomechanical back-action rate ($\gammaOMO$) is also approximately equal to the scattering rate from the $0$th sideband to the $1$st Stokes sideband ($\gamma_{0,1}$) in the linearized regime, which for $\Delta=-\omegamO$ blue detuned pumping yields $|\gammaOMO|\approx\gamma_{0,1}\approx 4G^2/\kappa$.  As this is the case, in the main text we don't differentiate between back-action damping and Stokes scattering rate, and $\ncavO$ is a simple placeholder for $P_{\text{in}}$ in all of our plots.  

\end{document}